\documentclass[aps,prc,superscriptaddress,twoside,twocolumn,%
                nofootinbib,showpacs]{revtex4-1}
\usepackage{amsmath,amssymb}
\usepackage{graphicx}
\usepackage{braket}

\newcommand*{\fs}[1]{#1\!\!\!/}
\newcommand*{\tJ}{\tilde{J}}
\newcommand*{\tkappa}{\tilde{\kappa}}
\newcommand*{\lon}{{\textsc{l}}}
\newcommand*{\tra}{{\textsc{t}}}
\newcommand*{\ba}{{\text{\tiny b}}}
\newcommand*{\KR}{{\textsc{kr}}}
\newcommand*{\IC}{{\text{int}}}

\allowdisplaybreaks

\begin{document}

\title{Dressing the electromagnetic nucleon current}

\author{H.~Haberzettl}
 \email{helmut.haberzettl@gwu.edu}
 \affiliation{\mbox{Center for Nuclear Studies, Department of Physics, The George Washington University, Washington, DC 20052, USA}}
\author{F.~Huang}
 \email{huang@physast.uga.edu}
 \affiliation{Department of Physics and Astronomy, University of Georgia, Athens, GA 30602, USA}
\author{K.~Nakayama}
 \email{nakayama@uga.edu}
 \affiliation{Department of Physics and Astronomy, University of Georgia, Athens, GA 30602, USA}
 \affiliation{\mbox{Institut f{\"u}r Kernphysik and J\"ulich Center for Hadron Physics, Forschungszentrum J{\"u}lich, 52425 J{\"u}lich, Germany}}

\date{10 March 2011}

\begin{abstract}
A field-theory-based approach to pion photoproduction off the nucleon is used
to derive a microscopically consistent formulation of the fully dressed
electromagnetic nucleon current in an effective Lagrangian formalism. It is
shown how the rigorous implementation of local gauge invariance at all levels
of the reaction dynamics provides equations that lend themselves to practically
manageable truncations of the underlying nonlinearities of the problem. The
requirement of consistency also suggests a novel way of treating the pion
photoproduction problem. Guided by a phenomenological implementation of gauge
invariance for the truncated equations that has proved successful for pion
photoproduction, an expression for the fully dressed nucleon current is given
that satisfies the Ward-Takahashi identity for a fully dressed nucleon
propagator as a matter of course. Possible applications include meson photo-
and electroproduction processes, bremsstrahlung, Compton scattering, and $ee'$
processes off nucleons.
\end{abstract}

\pacs{13.40.-f, 13.60.Le, 25.20.Lj, 13.60.Fz}

\maketitle

\section{Introduction}  \label{sec:introduction}


The electromagnetic interaction provides the cleanest probe of hadronic systems
available to experimentalists. Many experimental facilities, such as JLab,
MAMI, ELSA, SPring-8, GRAAL, and many others around the world, therefore, use
reactions employing real or virtual photons to gain information about the
internal dynamics of hadronic systems. (For a recent review, see
Ref.~\cite{Klempt2010}.)

However, while we understand the electromagnetic interaction perfectly well at
the elementary level, its applications in actual experiments do not concern
elementary particles but composite systems of elementary particles that
describe the internal structures of the baryonic or mesonic systems that take
part in the experiments. At intermediate energies, for baryons, in particular,
there usually is no need to invoke quark degrees of freedom to understand their
internal structures since the internal dynamics of baryons can be described
very well in terms of baryonic and mesonic degrees of freedom.

One very successful, quite fundamental way of dealing with these degrees of
freedom is the effective field-theory framework of chiral perturbation
theory~\cite{ChPT}. However, in view of its perturbative nature, this cannot be
easily extended to energy regions too far away from threshold. At higher
energies, one usually must rely on effective Lagrangian formulations that offer
a more direct avenue to the actual meson and baryon degrees of freedom that
manifest themselves in the experiments.

It is important, therefore, to understand the nature of the electromagnetic
interaction with mesons and baryons in a more detailed picture. One of the most
important and most basic systems in this respect is the nucleon itself.

The matrix element of the electromagnetic current operator $J^\mu$ of the
nucleon between on-shell nucleon spinors is given by\footnote{The term
 proportional to the photon four-momentum $k^\mu$ is often left out here
 because it does not contribute for transverse photon states. However, this is
 the correct, general way of writing this on-shell current since the $k^2$
 dependence for virtual photons can only occur in manifestly transverse
 contributions. In this respect, see also the general discussion on the
 structure of the nucleon current in Ref.~\cite{KPS2001}.}
\begin{align}
\bar{u}J^\mu u &=  \bar{u}(p') \bigg[e \delta_N\,\gamma^\mu +e
\delta_N\left(\gamma^\mu k^2-k^\mu \fs{k}\right) \frac{F_1(k^2)-1}{k^2}
\nonumber\\
&\mbox{}\qquad\qquad
+  e\frac{i\sigma^{\mu\nu}k_\nu}{2m} \kappa_N F_2(k^2) \bigg] u(p)~,
\label{eq:JmuDirPau}
\end{align}
where $e$ is the fundamental charge unit, $\delta_N$ is 1 or 0 for proton or
neutron, respectively, and $\kappa_N$ is the nucleon's anomalous magnetic
moment; $m$ is the physical nucleon mass (which here is related to the incoming
and outgoing nucleon four-momenta by $p^2=p'^2=m^2$). The (scalar) Dirac and
Pauli form factors, $F_1$ and $F_2$, respectively, are functions of the squared
photon four-momentum $k=p'-p$, normalized here such that $F_1(0)=F_2(0)=1$. The
expression appearing within the square brackets, with two independent
coefficient functions, $F_1$ and $F_2$, is the most general expression for the
current $J^\mu$ for on-shell nucleons. As such, therefore, it appears only in
physical processes involving virtual photons, like electron scattering off the
nucleon, for example. While it is well known~\cite{Bincer1960} that any
physical mechanism involving off-shell nucleons, in general, (after having
applied all available symmetry constraints) requires an expansion of the
current operator in terms of six independent form factors, the simplified
expression~(\ref{eq:JmuDirPau}) nevertheless remains the current
parametrization of choice for many if not most descriptions of photoprocesses
within effective Lagrangian approaches irrespective of whether the photon is
real or the incoming and outgoing nucleons are on-shell. \textit{A priori}, of
course, it is not clear how much of the dynamics of the full electromagnetic
coupling to the nucleon is ignored by such a simplified approach.

It is the purpose of the present work to provide a more detailed description of
the electromagnetic nucleon current $J^\mu$. We will start from a comprehensive
field theory~\cite{Haberzettl1997} that utilizes baryon and meson degrees of
freedom to describe pion-nucleon scattering and that also provides --- via its
description of the dressed nucleon propagator --- an avenue to the detailed
dynamics of the nucleon's electromagnetic interaction. The full formalism is a
very complex and nonlinear Dyson-Schwinger-type approach and, as such,
therefore, not easily implemented in practical applications. We will show here
how this formalism can be reformulated equivalently in a manner that makes it
directly amenable to physically motivated approximation schemes, thus rendering
the approach practically manageable. Of decisive importance in this respect
will be the fact that the internal dressing effects of the nucleon current and
the dynamics of pion photoproduction are very closely related.

The paper is organized as follows. In Sec.~\ref{sec:current1},  concentrating
on contributions due to pions, nucleons, and photons only, we introduce some
basic facts needed for the description of the dressed nucleon current $J^\mu$.
In doing so, we follow the corresponding field-theory formulation of
Haberzettl~\cite{Haberzettl1997}. In particular, we discuss the structure of
the unique minimal current~\cite{BallChiu1980} that provides the current's
Ward-Takahashi identity~\cite{WTI,WeinbergI}. It is argued that the internal
structure of $J^\mu$ is very closely related to pion photoproduction and we
therefore revisit this problem in Sec.~\ref{sec:photoproduction} where we
extend the approach of Haberzettl, Nakayama, and Krewald~\cite{Haberzettl2006}
to make the truncated formalism gauge invariant in a manner that is
microscopically consistent with the dressing mechanisms of the nucleon current
that are provided in the subsequent Sec.~\ref{sec:current2}. Finally,
Sec.~\ref{sec:summary} provides a summarizing assessment, including a
discussion of possible applications. Throughout the presentation of our
formalism, we discuss possible approximations to render the complex
nonlinearity of the resulting equations manageable in practice.

\section{Nucleon Current: Basic Considerations}  \label{sec:current1}


\begin{figure*}[t!]\centering
  \includegraphics[width=.8\textwidth,clip=]{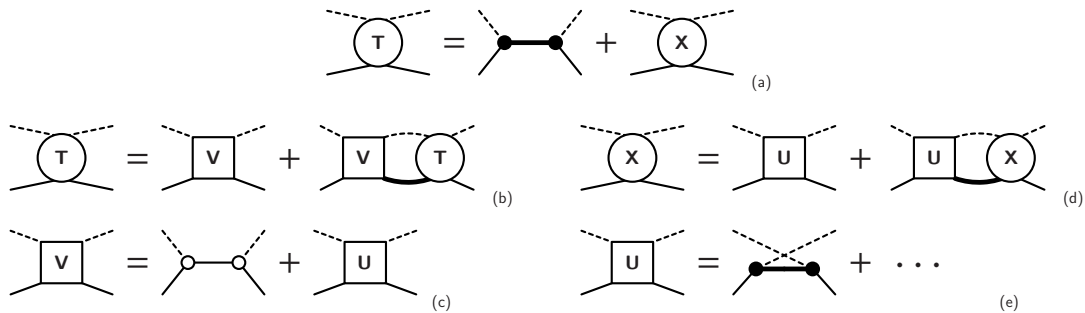}
  \caption{\label{fig:TXVU}%
  Generic structure of the pion-nucleon $T$ matrix employing pions and nucleons
  as the only hadronic degrees of freedom~\cite{Haberzettl1997}. (a) Splitting
  of $T$ into $s$-channel pole part and non-pole $X$. (b) Bethe-Salpeter
  integral equation for $T$, with (c) the driving term $V$ according to
  Eq.(\ref{eq:VPOT}). (d) Bethe-Salpeter integral equation for non-pole $X$,
  with (e) non-pole driving term $U$. Dressed vertices are solid circles;
  undressed ones open circles. Dressed (internal) nucleons are shown as thick
  lines; undressed ones as thin lines; pions are shown as dashed lines. Note
  that the $s$-channel pole term in the driving term $V$ is bare [because it
  gets dressed by the equation (b) itself] whereas, in the full theory, all
  mechanisms in the non-pole $U$ are fully dressed via Dyson-Schwinger-type
  mechanisms, as depicted in Fig.~\ref{fig:hadrondressing}.}
\end{figure*}
%

\begin{figure}[t!]\centering
\includegraphics[width=.7\columnwidth,clip=]{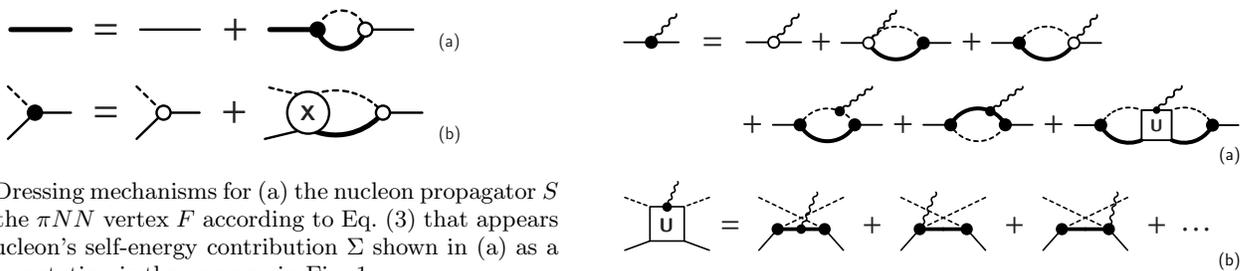}
  \caption{\label{fig:hadrondressing}%
  Dressing mechanisms for (a) the nucleon propagator $S$ and (b) the $\pi NN$
  vertex $F$ according to Eq.~(\ref{eq:Fvertex}) that appears in the nucleon's
  self-energy contribution $\Sigma$ shown in (a) as a loop. The notation is the
  same as in Fig.~\ref{fig:TXVU}.}
\end{figure}
%

The generic structure of the electromagnetic current $J^\mu$ of the nucleon can
be determined in a formulation that involves pions and nucleons and photons
only. Any additional hadronic degrees of freedom will only complicate the
situation, but will not add any qualitatively new structure to $J^\mu$; in
other words, they will not add anything of substance to the
discussion.\footnote{Of course, in an actual application of the present
  formalism, one must include
  all relevant particle degrees of freedom.} Using these degrees of freedom, the field-theory
approach of Haberzettl~\cite{Haberzettl1997} provides an expression for the
current based on a Lorentz-covariant effective Lagrangian formalism. Rather
than recapitulating all of the details of Ref.~\cite{Haberzettl1997}, we
summarize the result given there in several diagrams.

To define the dressed nucleon current $J^\mu$, we need the dressed nucleon
propagator $S$ which is obtained from the $T$ matrix for $\pi N$ scattering.
Figure~\ref{fig:TXVU} shows the structure of this $T$ matrix;
Fig.~\ref{fig:TXVU}(a), in particular, depicts the splitting of the full
amplitude $T$ into its $s$-channel pole part and its non-pole part $X$, i.e.,
\cite{Haberzettl1997}
\begin{equation}
T = \ket{F} S \bra{F} +X~,
\label{eq:TXsplit}
\end{equation}
relevant for some of the present considerations.\footnote{We follow here the
  notation of Ref.~\cite{Haberzettl1997}, i.e., we do not use the usual
  notation of $T^{\text{P}}$ and $T^{\text{NP}}$ for the pole and non-pole
  contributions of $T$, respectively, because the corresponding indices tend to
  clutter up the equations. For the same reason, we use $U$ instead of
  $V^{\text{NP}}$ for the driving term  of the non-pole Bethe-Salpeter equation
  (\ref{eq:BS_X}). Furthermore, as in Ref.~\cite{Haberzettl1997}, the bra and
  ket notation is used here simply as a quick way to see whether a vertex $F$
  describes $N\to \pi N$ --- which would be $\ket{F}$ --- or $\pi N\to N$,
  which is written as $\bra{F}$. This avoids the excessive use of adjoint
  daggers ($\dagger$) and makes the equations easier to read. The bras and kets
  are not to be misconstrued as Hilbert-space vectors.}
The first term here contains the nucleon propagator $S$ that provides the
$s$-channel pole. The $F$s are the fully dressed $\pi NN$ vertices related to
the bare vertex $f$  by
\begin{equation}
  \ket{F}=\ket{f} + XG_0 \ket{f}~,
  \label{eq:Fvertex}
\end{equation}
which is part of the nonlinear Dyson-Schwinger-type equations shown in
Fig.~\ref{fig:hadrondressing}. Both $T$ and $X$ are obtained as solutions of
Bethe-Salpeter-type integral equations according to
\begin{equation}
T = V +V G_0 T~,
\label{eq:BS_T}
\end{equation}
and
\begin{equation}
X=U + UG_0 X~,
\label{eq:BS_X}
\end{equation}
as depicted in Figs.~\ref{fig:TXVU}(b) and \ref{fig:TXVU}(d), respectively. The
respective driving terms $V$ and $U$ differ by the bare $s$-channel diagram, as
shown in Figs.~\ref{fig:TXVU}(c) and \ref{fig:TXVU}(e), i.e.,
\begin{equation}
V = \ket{f} S_0 \bra{f} + U~,
\label{eq:VPOT}
\end{equation}
where $S_0$ stands for the bare nucleon propagator.
 $G_0$ in Eqs.~(\ref{eq:Fvertex})-(\ref{eq:BS_X}) describes the
intermediate propagation of free pion and nucleon states that share the same
total four-momentum of the process.

\begin{figure}[t!]\centering
  \includegraphics[width=.95\columnwidth,clip=]{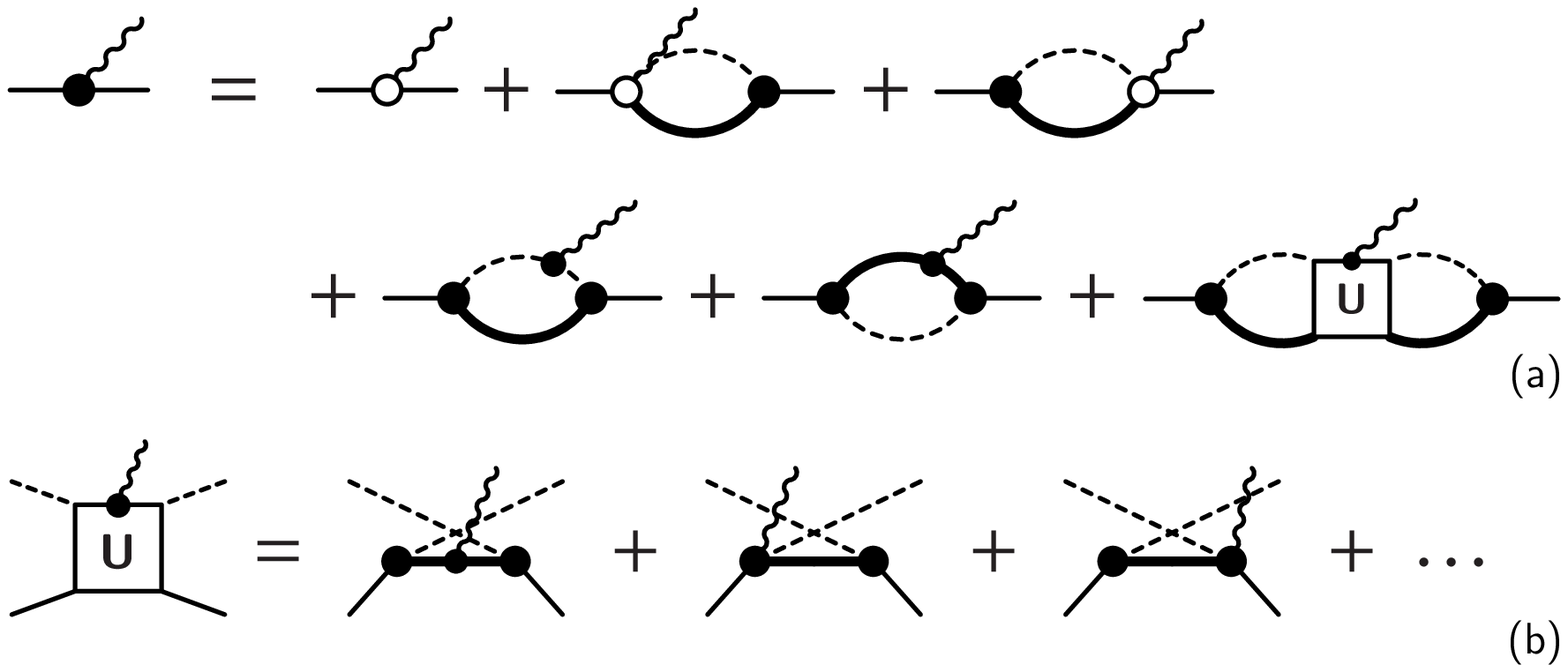}
  \caption{\label{fig:NCurrentFull}%
  (a) Structure of the full electromagnetic nucleon current $J^\mu$ employing
  nucleons and pions as the only hadronic degrees of
  freedom~\cite{Haberzettl1997}. The open circle of the first term on the
  right-hand side is the bare current $J^\mu_0$ and open-circle four-point
  functions in the next two diagrams depict the Kroll-Ruderman contact current.
  The last diagram subsumes the intermediate contributions of the interaction
  current $U^\mu$ arising from the photon interacting with the internal
  mechanisms of the one-nucleon irreducible (i.e., non-polar) $\pi N$
  interaction $U$. (b) The interaction current $U^\mu$; explicitly shown are
  only the lowest-order contributions that follow from the photon interacting
  with the $u$-channel Born term of $\pi N$ scattering [cf.\
  Fig.~\ref{fig:TXVU}(e)]. The $\gamma\pi NN$ four-point vertices of the last
  two diagrams subsume the interaction of the photon with the interior of the  fully dressed
  $\pi NN$ vertex (cf.\ last diagram in Fig.~\ref{fig:PionPhotoGeneric}).}
\end{figure}

\begin{figure}[t!]\centering
  \includegraphics[width=\columnwidth,clip=]{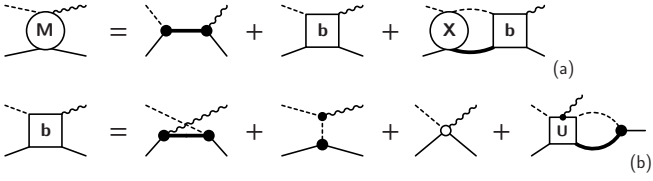}
  \caption{\label{fig:MCurrent}%
  (a) Pion photoproduction current $M^\mu$~\cite{Haberzettl1997}. The diagrams
  show the splitting of the production current $M^\mu$ into the $s$-channel
  pole term and the remaining one-nucleon irreducible contributions, with the
  final-state interaction mediated by the non-pole part $X$ of the pion-nucleon
  $T$ matrix. (b) Structure of the Born-type contribution $b^\mu$, as given in
  Eq.~(\ref{eq:fullBorn}). The four diagrams on the right-hand side depict, in
  the order given, the $u$- and $t$-channel contributions, the Kroll-Ruderman
  contact term, and the loop involving the $\pi N$ interaction current $U^\mu$
  of Fig.~\ref{fig:NCurrentFull}(b).}
\end{figure}

\begin{figure}[t!]
  \includegraphics[width=.85\columnwidth,clip=]{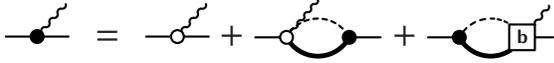}
  \caption{\label{fig:NCurrentPhoto}%
   Alternative depiction of the nucleon current $J^\mu$ employing the Born-type
   pion-production contribution $b^\mu$ from Fig.~\ref{fig:MCurrent}(b).}
\end{figure}

The fully dressed electromagnetic nucleon current $J^\mu$ derived in
Ref.~\cite{Haberzettl1997} is shown in Fig.~\ref{fig:NCurrentFull}. Formally,
it results from applying the gauge-derivative procedure~\cite{Haberzettl1997}
to the dressed nucleon propagator $S$, however, it can be understood very
simply as attaching a photon line to the propagator diagrams in
Fig.~\ref{fig:hadrondressing}(a) in all possible ways. To further understand
the details of this structure, we mention that one of the simplest physical
manifestations of the nucleon current occurs in the pion photoproduction
process off the nucleon (shown in Fig.~\ref{fig:MCurrent}) because here the
nucleon current provides one of the factors of the $s$-channel pole term (the
other being the hadronic  $\pi NN$ production vertex). It should not be
surprising, therefore, that much of the detailed internal structure of the
current can be understood by the same mechanisms that contribute to the pion
photoproduction amplitude $M^\mu$. Figure \ref{fig:NCurrentPhoto} shows that
all internal dynamics of the nucleon current $J^\mu$  shown in
Fig.~\ref{fig:NCurrentFull}(a) may be represented equivalently in terms of
loops over one-nucleon irreducible contributions to the pion photoproduction,
with the exception of one loop involving the Kroll-Ruderman
current~\cite{Haberzettl1997}. This close relationship of the dressing
mechanisms of the nucleon current forms the basis of the results presented
below.

%

%

%

\subsection{Gauge invariance}


The dressed current $J^\mu$ must satisfy  gauge invariance; hence, it must obey
the Ward-Takahashi identity (WTI)~\cite{WTI,WeinbergI},
\begin{equation}
  k_\mu J^\mu(p',p) = S^{-1}(p') Q_N - Q_N S^{-1}(p)~,
  \label{eq:WTI}
\end{equation}
where $p$ and $p'$ are the incoming and outgoing nucleon four-momenta,
respectively, and $k=p'-p$ is the (incoming) photon momentum; $Q_N$ is the
nucleon's charge operator. This \textit{off-shell constraint} ensures  a
conserved current for nucleons that are on-shell, i.e., when $p'^2=p^2=m^2$

Without lack of generality, we may write the nucleon current as
\begin{equation}
  J^\mu(p',p) =  J_s^\mu(p',p) + T^\mu(p',p)~,
\end{equation}
where $J_s^\mu$ is the \textit{minimal} current that satisfies the WTI
(\ref{eq:WTI}), i.e.,
\begin{equation}
  k_\mu J_s^\mu(p',p) = S^{-1}(p') Q_N - Q_N S^{-1}(p)~,
  \label{eq:WTIs}
\end{equation}
and $T^\mu$ thus is the transverse remainder, with
\begin{equation}
  k_\mu T^\mu(p',p) = 0~.
\end{equation}
By construction, this transversality must be manifest globally and it is not
subject to any particular kinematic or dynamic restrictions.

\subsection{Minimal nucleon current}


Let us write the dressed propagator for a nucleon with four-momentum $p$ in a
generic manner as
\begin{equation}
  S(p) = \frac{1}{\fs{p}\,A(p^2)-m\,B(p^2)}~,
  \label{eq:SFstructure}
\end{equation}
where $A(p^2)$ and $B(p^2)$ are the two independent scalar dressing functions
constrained by the residue conditions
\begin{subequations}
\begin{equation}
  A(m^2)=B(m^2)
\end{equation}
and
\begin{equation}
A(m^2)+2m^2\frac{d\left[A(p^2)-B(p^2)\right]}{dp^2}\bigg|_{p^2=m^2}=1~.
 \label{eq:residue}
\end{equation}
\end{subequations}
From the residue condition alone, one cannot in general conclude that
$A(m^2)=B(m^2)=1$; in the structureless case, however, we have $A\equiv B\equiv
1$. (Note, however, that even though there are no explicit $p^2$-dependent
dressing functions in the latter case, implicit dressing effects are present
nevertheless owing to the fact that the mass $m$ is the physical mass.)

Following Ball and Chiu~\cite{BallChiu1980}, the minimal nucleon current that
satisfies the WTI (\ref{eq:WTIs}) is given by
\begin{align}
J_s^\mu(p',p) &= (p'+p)^\mu \frac{S^{-1}(p')Q_N-Q_NS^{-1}(p)}{p'^2-p^2}
\nonumber\\
&\mbox{}\quad
+\left[\gamma^\mu -
\frac{(p'+p)^\mu}{p'^2-p^2}\fs{k}\right]Q_N\frac{A(p'^2)+A(p^2)}{2}~.
 \label{eq:JsDefined}
\end{align}
The first term here on the right-hand side is sufficient to produce the WTI,
but the second part (which is transverse) is necessary to fully cancel the
$1/(p'^2-p^2)$ singularity, as can be seen explicitly by recasting $J_s^\mu$ in
the equivalent form
\begin{align}
  J_s^\mu(p',p)
  & = \gamma^\mu Q_N \frac{ A(p'^2)+A(p^2) }{2}
  \nonumber\\
    &\quad\mbox{}+
   (p'+p)^\mu Q_N
   \bigg[\frac{\fs{p}'+\fs{p}}{2}\frac{A(p'^2)-A(p^2)}{p'^2-p^2}
   \nonumber\\
   &\mbox{} \qquad\qquad\qquad -m\frac{B(p'^2)-B(p^2)}{p'^2-p^2}\bigg]~.
   \label{eq:JsNonsingular}
\end{align}
In fact, $J^\mu_s$ is the \textit{unique} current that satisfies the WTI and
also is nonsingular and symmetric in $p'$ and $p$. Moreover, as can be seen
from (\ref{eq:JsNonsingular}), for structureless nucleons, this reduces to the
usual $\gamma^\mu$ Dirac current. And, invoking the generalized Gordon identity
\begin{equation}
  (p'+p)^\mu = -i \sigma^{\mu\nu}k_\nu +\fs{p}'\gamma^\mu +\gamma^\mu \fs{p}~,
  \label{eq:gGordon}
\end{equation}
the on-shell matrix element of $J^\mu_s$ is easily found as
\begin{align}
\bar{u}J_s^\mu u &=  \bar{u}(p') e\delta_N \bigg\{\gamma^\mu
+ i \frac{\sigma^{\mu\nu}k_\nu}{2m} \left[A(m^2)-1\right] \bigg\} u(p)~.
\label{eq:JsOnShell}
\end{align}
Note that there is no $k^2$ dependence here, i.e., this result does not depend
on whether the photon is real or virtual. This is consistent with the fact that
minimal currents that satisfy the WTI as a rule cannot depend on the photon
four-momentum since such a dependence \textit{always} sits in transverse
contributions~\cite{KPS2001}. We point out in this context that the
$\sigma^{\mu\nu}k_\nu$ contribution here must not be confused with the usual
Pauli current, i.e., its coefficient is \textit{not} directly related  to the
anomalous magnetic moment of the nucleon (which should be obvious because the
entire current $J^\mu_s$ vanishes for the neutron).

\subsubsection*{Minimal current taken half on-shell}

Of particular interest for many applications is to consider the half-on-shell
reduction of the current.  We shall do so here for an incoming on-shell nucleon
interacting with a photon followed by the subsequent propagation of an
off-shell nucleon, but the following considerations can be readily translated
into describing the reversed situation where the outgoing nucleon is on-shell.
Thus, half-on-shell, with an incoming nucleon spinor $u(p)$ on the right and an
outgoing propagator $S(p+k)$ on the left, using~(\ref{eq:gGordon}), this
results in
\begin{align}
SJ^\mu_s u &\equiv S(p+k)J^\mu_s(p+k,p)u(p)
\nonumber\\
&=
\left(\frac{1}{\fs{p}+\fs{k}-m}j^\mu_1
  + \frac{2m}{s-m^2}j^\mu_2\right)Q_N \, u(p)~,
  \label{eq:JsRonshellj}
\end{align}
 where $s=(p+k)^2$ and $p^2=m^2$. The (dimensionless) auxiliary currents are given by
\begin{subequations}\label{eq:auxcurrent}
\begin{align}
j^\mu_1 &= \gamma^\mu\,\left(1-\kappa_1\right)
  + \frac{i\sigma^{\mu\nu}k_\nu}{2m} \kappa_1
  \label{eq:auxj1}
\\
\intertext{and}
j^\mu_2 &= \frac{(2p+k)^\mu}{2m} \kappa_1+ \frac{i\sigma^{\mu\nu}k_\nu}{2m} \kappa_2~,
  \label{eq:auxj2}
\end{align}
\end{subequations}
with (dimensionless) \textit{independent} coefficient functions
\begin{subequations}\label{eq:kappadefined}
\begin{align}
  \kappa_1 &=
   m^2\frac{\left[B(s)-A(s)\right]\left[A(s)+A(m^2)\right]}{sA^2(s)-m^2B^2(s)}
  \\
  \intertext{and}
  \kappa_2 &=\frac{A(m^2)-A(s)}{2A(s)} + \frac{A(s)+B(s)}{2A(s)}\kappa_1~.
\end{align}
\end{subequations}
The on-shell values at $s=m^2$ of both coefficients are identical, i.e.,
\begin{equation}
  \kappa_1(m^2)=\kappa_2(m^2) =A(m^2)-1~.
\end{equation}
This means they both vanish in the structureless limit, thus leaving in
(\ref{eq:JsRonshellj}) only the usual $\gamma^\mu$ Dirac current together with
a structureless propagator. All effects of the dressing thus reside in the
terms that depend on the $\kappa_i$ ($i=1,2$) whose overall contributions are
easily seen to be transverse.

We emphasize that Eq.~(\ref{eq:JsRonshellj}) is exact and that it includes
\textit{all} possible dressing mechanisms. Its four-divergence, in particular,
is given by
\begin{equation}
  k_\mu \,S(p+k) J^\mu_s(p+k,p)\,u(p) = Q_N \,u(p)
\end{equation}
and the resulting expression does not involve any dressing effects whatsoever.
Any approximation, therefore, that only involves the coefficient functions
$\kappa_i$ will have no bearing on the gauge-invariance contribution of any
term containing (\ref{eq:JsRonshellj}). This is of direct and immediate
relevance for the treatment of pion photoproduction presented in the following.

\section{Pion photoproduction revisited}\label{sec:photoproduction}


In order to see how the gauge-invariant minimal current contribution $J^\mu_s$
of Eq.~(\ref{eq:JsDefined}) can be utilized for a practically useful
description of the full nucleon current $J^\mu$, we need to revisit the
photoproduction of the pion because, as alluded to above, the internal dynamics
of the current is closely related to mechanisms found in this production
process, as seen in Fig.~\ref{fig:NCurrentPhoto}.

\begin{figure}[t!]\centering
  \includegraphics[width=.9\columnwidth,clip=]{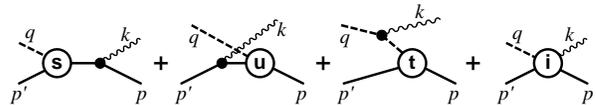}
  \caption{\label{fig:PionPhotoGeneric}%
  Generic pion photoproduction diagrams $\gamma +N \to \pi +N$, with the $s$-,
  $u$-, and $t$-channel pole diagrams $M^\mu_s$, $M^\mu_u$, and $M^\mu_t$,
  respectively, and the contact-type interaction current $M^\mu_\IC$ that
  subsumes all final-state interaction [cf.\ Eq.~(\ref{eq:Mint})]. The labels
  (s, u, t) at the $\pi NN$ vertices allude to the usual Mandelstam variables
  $s$, $u$, and $t$ that describe the respective kinematic situations. The
  four-momenta here are the ones used in Eqs.~(\ref{eq:WTIcontact}) and
  (\ref{eq:WTIcontactAmended}).}
\end{figure}

Following Ref.~\cite{Haberzettl1997}, we start by writing the production
current $M^\mu$ as
\begin{equation}
  M^\mu = \ket{F} S J^\mu + b^\mu + XG_0 b^\mu
  \label{eq:MmuX}
\end{equation}
which is a self-evident transcription of Fig.~\ref{fig:MCurrent}(a). The first
term on the right-hand side, the $s$-channel current $M^\mu_s$, contains the
nucleon pole and, in the last term, $X$ provides the $\pi N$ final-state
interaction (FSI) of the production current. The current $b^\mu$ subsuming the
Born-type mechanisms may be written as
\begin{equation}
  b^\mu = M^\mu_u +M^\mu_t + m^\mu_\KR + U^\mu G_0 \ket{F}
  \label{eq:fullBorn}
\end{equation}
which describes the four terms appearing on the right-hand side of
Fig.~\ref{fig:MCurrent}(b), which are, respectively, the $u$- and $t$-channel
currents, the Kroll-Ruderman current~\cite{Kroll1954} and the loop integration
involving the $\pi N$ interaction current $U^\mu$ of
Fig.~\ref{fig:NCurrentFull}(b). Generically, the overall structure of $M^\mu$
is presented in Fig.~\ref{fig:PionPhotoGeneric}, with the first three diagrams
containing the respective $s$-, $u$-, and $t$-channel pole contributions, and
everything else
--- including the FSI contributions --- being subsumed in the non-polar
four-point interaction current $M^\mu_\IC$.

%

As discussed in Ref.~\cite{Haberzettl1997}, the five-point $\pi N$ interaction
current $U^\mu$ contains very complicated mechanisms that cannot, in general,
easily be calculated exactly. Any approximations made necessary by
considerations of practicality, however, need to maintain the gauge invariance
of the amplitude $M^\mu$. The prescription put forward in
Ref.~\cite{Haberzettl2006} generalizes the basic procedure of
Ref.~\cite{Haberzettl1998} to allow the inclusion of the full FSI contribution
appearing in Eq.~(\ref{eq:MmuX}) in terms of $X$. This procedure based on the
generalized Ward-Takahashi identity for the production
current~\cite{Haberzettl1997,Kazes1959} is not unique, of course, since the
generalized WTI does not constrain transverse current contributions. We will
exploit this ambiguity here and provide an alternative gauge-invariant
approximation of the amplitude $M^\mu$ that will turn out to be very useful in
describing the mechanisms of the nucleon current $J^\mu$.

To this end, we split the Born current $b^\mu$ into its longitudinal and
transverse parts,
\begin{equation}
  b^\mu=b^\mu_\lon +b^\mu_\tra~,
\end{equation}
as indicated by the respective indices L and T, and write the FSI term of
(\ref{eq:MmuX}) equivalently as
\begin{align}
  XG_0 b^\mu &= XG_0 b^\mu_\lon + XG_0\left[\left(M^\mu_u +M^\mu_t\right)_\tra+T^\mu_X\right]
  \nonumber\\
  &\mbox{}\qquad
  + XG_0\left[\left(m^\mu_\KR +U^\mu G_0 \ket{F}\right)_\tra-T^\mu_X\right]~.
\end{align}
We have introduced here an as yet undetermined \textit{transverse} current
$T^\mu_X$ that cancels out in the last two terms; its choice, therefore, is of
no consequence for the full formalism. Inserting this into
(\ref{fig:NCurrentPhoto}), we may write
\begin{align}
  M^\mu &= \ket{F} S J^\mu  +M^\mu_u +M^\mu_t+M^\mu_c
  \nonumber\\
  &\mbox{}\qquad
  + XG_0\left[\big(M^\mu_u +M^\mu_t\big)_\tra+T^\mu_X\right]~,
  \label{eq:Mmusplit}
\end{align}
where
\begin{align}
  M^\mu_c &= (1+XG_0)\big(m^\mu_\KR +U^\mu G_0 \ket{F}\big)
  \nonumber\\
  &\mbox{}\qquad
  + XG_0\big(M^\mu_u +M^\mu_t\big)_\lon -XG_0 T_X^\mu~.
  \label{eq:Mc}
\end{align}
The last two terms in (\ref{eq:Mmusplit}) describe the content of the
interaction current (cf.\ the last diagram in Fig.~\ref{fig:PionPhotoGeneric}),
i.e.,
\begin{equation}
  M^\mu_\IC = M^\mu_c
  + XG_0\left[\big(M^\mu_u +M^\mu_t\big)_\tra+T^\mu_X\right]~,
  \label{eq:Mint}
\end{equation}
and thus both $M^\mu_\IC$ and $M^\mu_c$ must obey identical constraints to
render $M^\mu$ gauge invariant. Specifically, their four-divergences must
satisfy~\cite{Haberzettl1997,Haberzettl2006}
\begin{equation}
  k_\mu M^\mu_\IC = k_\mu M^\mu_c=
  -F_s e_i +F_u e_f +F_t e_\pi~,
  \label{eq:GIcondition}
\end{equation}
where $k$ is the incoming photon four-momentum and the $F_x$ describe the $\pi
NN$ vertices (including coupling operators) in the kinematic situations
corresponding to the Mandelstam variables $x=s,u,t$, as shown in
Fig.~\ref{fig:PionPhotoGeneric}; $e_i$, $e_f$, and $e_\pi$ comprise the
combined isospin and charge-operator factors for the initial and final
nucleons, and for the pion, respectively, of the vertices that provide charge
conservation in the form $e_i=e_f+e_\pi$. We emphasize that the
gauge-invariance condition (\ref{eq:GIcondition}) is an \textit{off-shell}
constraint that must always be true, i.e., it is not restricted to special
kinematic or dynamic situations.

\subsection{\boldmath Making $M^\mu$ gauge invariant}\label{sec:MakeGIP}


The expressions for the current $J^\mu$ derived here are exact presuming that
all contributions are calculated in the self-consistent manner prescribed by
the (nonlinear) Dyson-Schwinger-type equations summarized in
Figs.~\ref{fig:TXVU} through \ref{fig:NCurrentPhoto}. In practice,
unfortunately, this is quite out of the question in view of the enormous
complexity of the nonlinear problem. Since any truncation of the full formalism
is very likely to result in the violation of gauge invariance, one then needs
to introduce gauge-invariance preserving (GIP) procedures to restore it. One of
the big advantages of the formulation given here is that as long as the
electromagnetic nucleon and pion currents satisfy their individual WTIs, any
truncations necessitated by practicality can always be expressed in terms of
approximations of $M^\mu_c$. And as long as these approximations are chosen to
satisfy the interaction-current condition (\ref{eq:GIcondition}), gauge
invariance will be preserved as a matter of course.

As discussed in Ref.~\cite{Haberzettl2006}, there are various levels of
sophistication at which the constraint (\ref{eq:GIcondition}) can be
implemented, depending on how much of the detailed dynamics of $M^\mu_c$ can be
incorporated in a particular application. For the present purpose, it is
sufficient to point out that the procedures to do so are well defined and
straightforward to implement.

At the simplest level, the $\pi NN$ vertices are described by phenomenological
form factors, and we may then approximate all of $M^\mu_c$ by the
phenomenological GIP current~\cite{Haberzettl1998,Haberzettl2006}
\begin{align}
  M^\mu_c \to M^\mu_c& =
  - (1-\lambda) g \frac{\gamma_5\gamma^\mu}{2m} \tilde{F}_t e_\pi
  \nonumber\\[2ex]
  &\mbox{}\qquad
  - G_\lambda \Bigg[ e_i \frac{(2p+k)^\mu}{s-p^2}\left(\tilde{F}_s -\hat{F}\right)
  \nonumber\\[2ex]
  &\mbox{}\qquad\qquad
  +  e_f \frac{(2p'-k)^\mu}{u-p'^2} \left( \tilde{F}_u-\hat{F}\right)
  \nonumber\\[2ex]
  &\mbox{}\qquad\qquad
  +  e_\pi\frac{(2q-k)^\mu}{t-q^2}\left(\tilde{F}_t- \hat{F}\right)\Bigg]~,
  \label{eq:WTIcontact}
\end{align}
where the momenta shown in Fig.~\ref{fig:PionPhotoGeneric} are being used. The
first term here provides a dressed version of the Kroll-Ruderman
current\footnote{\label{foot:KR}%
  Note that $\tilde{F}_t=1+(\tilde{F}_t-1)$,
  i.e., the original bare Kroll-Ruderman term $m^\mu_\KR$ survives in this GIP
  current and the phenomenological dressing comes via the additional $\tilde{F}_t-1$
  contribution.}
and the other three terms supply the gauge-invariance corrections for the $s$-,
$u$, and $t$-channel contributions. The vertex functions with tilde have been
stripped of their isospin factors and coupling operators, i.e., $F_s Q_N =
G_\lambda e_i \tilde{F}_s$, etc. (see Ref.~\cite{Haberzettl2006} for more
technical details). The coupling operator is written as
\begin{equation}
G_\lambda=  g\, \gamma_5 \left[(1-\lambda) \frac{\fs{q}}{2m} + \lambda \right]~,
\label{eq:PSPVmixing}
\end{equation}
where $g$ is the coupling strength, $q$ is the outgoing pion four-momentum, and
$\lambda$ dials between pseudovector  ($\lambda=0$) and pseudoscalar
($\lambda=1$) coupling.\footnote{Note in this context that phenomenological
  form factors are intended to mock up the fully dressed vertex. Hence, even
  if one starts out with a fully chiral-symmetric pseudovector bare vertex, the
  dressed vertex, in general, would no longer be pure pseudovector. The ansatz
  (\ref{eq:PSPVmixing}) accounts for this fact in a phenomenological manner.}
The function $\hat{F}$ must be chosen such that any one of the three terms in
the square brackets remains finite if the corresponding denominators go to
zero. For specific choices of how to achieve this, see
Ref.~\cite{Haberzettl2006}. This phenomenological expression is then
non-singular and, moreover, it clearly satisfies the gauge-invariance condition
(\ref{eq:GIcondition}).

The procedure to preserve gauge invariance just described is the one put
forward in Ref.~\cite{Haberzettl2006}, with the specific choice $T_X^\mu=0$ for
the (undetermined) transverse current appearing in Eq.~(\ref{eq:Mmusplit}),
since it has no bearing on the gauge invariance. We emphasize that the only
property of the nucleon current $J^\mu$ itself that entered the considerations
so far was the usual Ward-Takahashi identity (\ref{eq:WTI}) for the nucleon and
its analog for the pion. In the following, we will show that taking into
account the details of the internal structure of the current $J^\mu$ as
depicted in Fig.~\ref{fig:NCurrentPhoto} will actually suggest a different
(i.e., non-zero) choice for $T^\mu_X$.

\subsection{\boldmath Choosing $T^\mu_X$}\label{sec:ChoosingTX}


While the choice of the (transverse) $T^\mu_X$ is irrelevant for gauge
invariance, it is clear, however, from Eq.~(\ref{eq:Mmusplit}), that any
particular choice for $T^\mu_X$ will have a definite impact on the overall
quality of the results if $M^\mu_c$ is subjected to approximations since
$M^\mu_c$ itself contains $T^\mu_X$ and therefore, in general, the results will
then no longer be independent of $T^\mu_X$.

\begin{figure}[t!]\centering
  \includegraphics[width=\columnwidth,clip=]{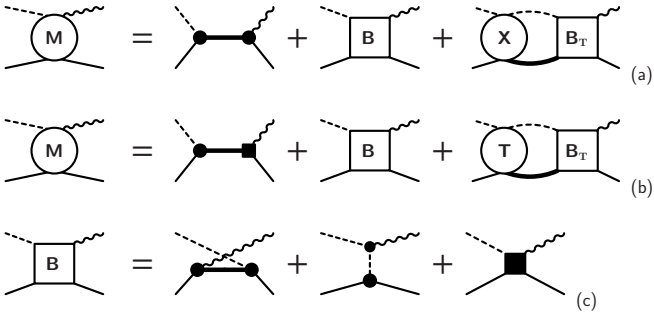}
  \caption{\label{fig:MwithBox}%
  Representations of the photoproduction current $M^\mu$ (equivalent to those
  of Fig.~\ref{fig:MCurrent}), with (a) loop integration over the non-pole
  amplitude $X$ [see Eq.~(\ref{eq:Mmusplit1})] and (b) loop integration over
  the full $\pi N$ $T$ matrix [see Eq.~(\ref{eq:MmuT})]. The details of the
  four-point box B are shown in (c), with the last diagram subsuming the
  mechanisms shown in Fig.~\ref{fig:McAll}. B$_\tra$ denotes the restriction of
  B to transverse contributions. The different nucleon-current contributions
  appearing in the $s$-channel terms of (a) and (b) are given in
  Fig.~\ref{fig:NcurrentAll}.}
\end{figure}
%

\begin{figure}[t!]\centering
  \includegraphics[width=\columnwidth,clip=]{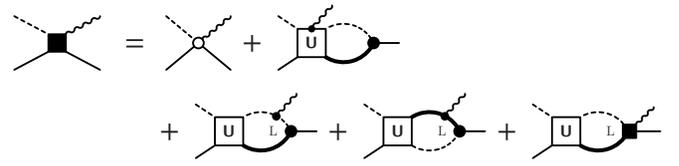}
  \caption{\label{fig:McAll}%
  Diagrammatic representation of Eq.~(\ref{eq:McEq2}), with the solid square
  four-point vertex depicting the contact-type current $M^\mu_c$. The first two
  diagrams on the right-hand side show $M^\mu_a$ of Eq.~(\ref{eq:MaDefined}).
  The indices L in the loops of the last three diagrams signify that only the
  longitudinal parts of the respective photon couplings are to be taken into
  account. These three diagrams, therefore, do not contribute for real
  photons.}
\end{figure}
%

\begin{figure*}[t!]\centering
  \includegraphics[width=.7\textwidth,clip=]{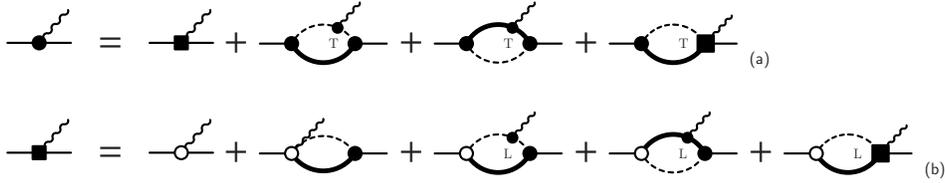}
  \caption{\label{fig:NcurrentAll}%
  (a) Dressed nucleon current $J^\mu$ according to Eq.~(\ref{eq:Ncurrent3}).
  The first term on the right-hand side (with a solid square vertex) depicts
  $\tJ^\mu_s$ which subsumes the mechanisms shown in part (b). Diagrams labeled
  T or L correspond to the transverse and longitudinal contributions given in
  Eqs.~(\ref{eq:Ncurrent3}) and (\ref{eq:JsExact}), respectively. The
  solid-square four-point vertices in the last diagrams of each line depict
  $M^\mu_c$ given in Fig.~\ref{fig:McAll}. The phenomenological approximations
  discussed in Sec.~\ref{sec:PhenDressing} affect the (square-shaped) three-
  and four-point vertices of $\tJ^\mu_s$ and $M^\mu_c$, respectively. }
\end{figure*}
%

We propose here to choose the undetermined transverse current as
\begin{equation}
T^\mu_X = \left(M^\mu_c\right)_\tra~.
\label{eq:MXchoice}
\end{equation}
The photoproduction amplitude then becomes
\begin{align}
  M^\mu &= \ket{F} S J^\mu  +M^\mu_u +M^\mu_t+M^\mu_c
  \nonumber\\
  &\mbox{}\qquad
  + XG_0\big(M^\mu_u +M^\mu_t +M^\mu_c\big)_\tra~,
  \label{eq:Mmusplit1}
\end{align}
which is shown in Fig.~\ref{fig:MwithBox}(a). $M^\mu_c$ in this expression
satisfies
\begin{align}
  M^\mu_c +XG_0\left(M^\mu_c\right)_\tra &= (1+XG_0) M^\mu_a
  \nonumber\\
  &\mbox{}\qquad+XG_0 \big(M^\mu_u +M^\mu_t\big)_\lon~,
  \label{eq:McEq1}
\end{align}
where
\begin{equation}
  M^\mu_a = m^\mu_\KR+U^\mu G_0\ket{F}
  \label{eq:MaDefined}
\end{equation}
subsumes the Kroll-Ruderman term and the loop integration over the  five-point
interaction current $U^\mu$. We may recast  (\ref{eq:McEq1}) in the form
\begin{equation}
M^\mu_c = M^\mu_a
+UG_0 \big(M^\mu_u +M^\mu_t+M^\mu_c\big)_\lon~,
  \label{eq:McEq2}
\end{equation}
which is a partial integral equation in the longitudinal part of $M^\mu_c$.
This equation is depicted in the diagrams of Fig.~\ref{fig:McAll}. We see here
that the choice (\ref{eq:MXchoice}) makes the transverse parts of $M^\mu_c$ and
$M^\mu_a$ identical, i.e.,
\begin{equation}
\left(M^\mu_c\right)_\tra \equiv \left(M^\mu_a\right)_\tra~.
\end{equation}
Since for real photons only the transverse parts of the currents contribute to
physical observables, this means that, from a practical point of view,
effectively any approximation of $M^\mu_c$ is a direct approximation of $U^\mu
G_0\ket{F}$,\footnote{Note that the Kroll-Ruderman term $m^\mu_\KR$ is
contained already in (\ref{eq:WTIcontact}); see Footnote~\ref{foot:KR}.} in
other words, of precisely that part of the production current $M^\mu$ which in
general cannot be calculated easily because of the complexity of its internal
dynamics~\cite{Haberzettl1997}. The choice (\ref{eq:MXchoice}), therefore,
provides an expression for the photoproduction current,
Eq.~(\ref{eq:Mmusplit1}), that is structurally very similar to the original
form (\ref{eq:MmuX}) and one that is readily amenable to approximations. Note,
in particular, that the fact that the explicit loop contribution in
(\ref{eq:Mmusplit1}) is only transverse is irrelevant for real photons since
they project out longitudinal contributions anyway, which effectively makes the
structure of (\ref{eq:Mmusplit1}) exactly the same as (\ref{eq:MmuX}) for such
photons.

\section{Dressing the Nucleon Current}\label{sec:current2}


Let us now turn back to the question of how to describe the dressing of the
nucleon current.

According to Fig.~\ref{fig:NCurrentFull}(a), the dressed current may be written
as~\cite{Haberzettl1997}
\begin{equation}
  J^\mu = J^\mu_\ba + \bra{F}G_0\left(M^\mu_u+M^\mu_t+M^\mu_a\right)~,
  \label{eq:Ncurrent1}
\end{equation}
where the  modified bare current $J^\mu_\ba$ is given by the first two diagrams
on the right-hand side of Fig.~\ref{fig:NCurrentFull},
\begin{equation}
J^\mu_\ba =  J_0^\mu + \braket{m_\KR^\mu|G_0|F}~,
\label{eq:JbDefined}
\end{equation}
where $J_0^\mu$ is the (true) bare current and the second term is the loop
containing the Kroll-Ruderman current $m_\KR^\mu$, however, with the pion
coming \textit{into} the contact vertex instead of going out.

The ingredients in the last term in Eq.~(\ref{eq:Ncurrent1}) are precisely
those that enter the photoproduction amplitude and we may, therefore, use
similar procedures to render them manageable in practical calculations.

Rewriting (\ref{eq:Ncurrent1}) equivalently in terms of $M^\mu_c$ of
Eq.~(\ref{eq:Mc}), by partially inverting the latter equation, we find
\begin{align}
  J^\mu &= J^\mu_\ba +\bra{f}G_0\left[\left(M^\mu_u+M^\mu_t\right)_\lon+M^\mu_c-T^\mu_X\right]
  \nonumber\\
  &\mbox{}\qquad
  +\bra{F}G_0 \left[\left(M^\mu_u+M^\mu_t\right)_\tra+T^\mu_X\right]~,
  \label{eq:Ncurrent2}
\end{align}
where the relationship (\ref{eq:Fvertex}) between dressed and undressed
vertices, $F$ and $f$, respectively, was used. In (\ref{eq:Ncurrent2}), the
$T^\mu_X$ dependence was left as in the original expressions to demonstrate
that the choice (\ref{eq:MXchoice}) is unique in the sense that it is the only
choice that produces an expression with a clear splitting of longitudinal and
transverse pieces, i.e., using  (\ref{eq:MXchoice}) we obtain
\begin{align}
J^\mu &= J^\mu_\ba + \bra{f}G_0\left(M^\mu_u+M^\mu_t+M^\mu_c\right)_\lon
\nonumber\\
&\mbox{}\qquad+ \bra{F}G_0 \left(M^\mu_u+M^\mu_t+M^\mu_c\right)_\tra~.
\end{align}
For later purposes, let us write this as
\begin{equation}
  J^\mu = \tJ^\mu_s + \bra{F}G_0 \left(M^\mu_u+M^\mu_t+M^\mu_c\right)_\tra~,
  \label{eq:Ncurrent3}
\end{equation}
where
\begin{equation}
\tJ^\mu_s = J^\mu_\ba +\bra{f}G_0\left(M^\mu_u+M^\mu_t+M^\mu_c\right)_\lon~.
\label{eq:JsExact}
\end{equation}
It is obvious, of course, that since $J^\mu$ and $\tJ^\mu_s$ differ only by
transverse pieces, their four-divergences coincide. Equations
(\ref{eq:Ncurrent3}) and (\ref{eq:JsExact}) are depicted in
Fig.~\ref{fig:NcurrentAll}.

We emphasize that expression~(\ref{eq:Ncurrent3}) for the dressed current is
exact as long as we do not employ any approximations.

\subsection{Phenomenological dressing}\label{sec:PhenDressing}


Apart from the modified bare contribution $J^\mu_\ba$, the result
(\ref{eq:Ncurrent3}) for the current provides a very suggestive even
distribution of the transverse and longitudinal contributions in a loop
integration over the fully dressed vertex for the former and a loop integration
over the bare vertex for the latter. Being related to the bare vertices, as
shown in Fig.~\ref{fig:NcurrentAll}(b), we would like to suggest that the
longitudinal contributions to $\tJ^\mu_s$ are minimal in the sense of the
minimal current $J^\mu_s$  of Eq.~(\ref{eq:JsDefined}), and that replacing
$\tJ^\mu_s$ by $J^\mu_s$ provides an excellent \textit{gauge-invariant}
approximation for this part of the full current. For $J^\mu$, instead of
Eq.~(\ref{eq:Ncurrent3}), we then write
\begin{equation}
  J^\mu \to J^\mu = J_s^\mu + \bra{F}G_0 \left(M^\mu_u+M^\mu_t+M^\mu_c\right)_\tra
  \label{eq:JmuPhen}
\end{equation}
which satisfies the usual WTI as a matter of course. Preliminary results
obtained for pion photoproduction show that (\ref{eq:JmuPhen}) is indeed an
excellent approximation~\cite{Huang2011}.

In addition, to ensure microscopic consistency among all associated reaction
mechanisms, we suggest to approximate $M^\mu_c$ similar to what was discussed
for photoproduction in Sec.~\ref{sec:MakeGIP}. For the present application to
the nucleon current, however, we also must make sure that the current
reproduces the anomalous magnetic moments for its on-shell matrix element [cf.\
Eq.~(\ref{eq:JmuDirPau})]. We, therefore, need to amend the approximation
(\ref{eq:WTIcontact}) according to
\begin{align}
  M^\mu_c \to M^\mu_c& =
  g e \gamma_5\frac{i\sigma^{\mu\nu}k_\nu}{4m^2} \tkappa_N
  - (1-\lambda) g \frac{\gamma_5\gamma^\mu}{2m} \tilde{F}_t e_\pi
  \nonumber\\[2ex]
  &\mbox{}\qquad
  - G_\lambda \Bigg[ e_i \frac{(2p+k)^\mu}{s-p^2}\left(\tilde{F}_s -\hat{F}\right)
  \nonumber\\[2ex]
  &\mbox{}\qquad\qquad
  +  e_f \frac{(2p'-k)^\mu}{u-p'^2} \left( \tilde{F}_u-\hat{F}\right)
  \nonumber\\[2ex]
  &\mbox{}\qquad\qquad
  +  e_\pi\frac{(2q-k)^\mu}{t-q^2}\left(\tilde{F}_t- \hat{F}\right)\Bigg]~,
  \label{eq:WTIcontactAmended}
\end{align}
with an additional transverse $\sigma^{\mu\nu}k_\nu$ current whose coefficient
$\tkappa_N$, in principle, needs to be fixed such that the on-shell matrix
elements of the current (\ref{eq:JmuPhen}) reproduce (\ref{eq:JmuDirPau}). The
factors in this term ensure that $\tkappa_N$ is dimensionless.  In practice,
the actual on-shell matrix elements of the nucleon current usually never enter
the calculations, and one may then use $\tkappa_N$ as an additional fit
parameter that accounts for the current being partially off-shell. As an
example of such a case, we discuss pion photoproduction in the following
section.

\subsection{Application to pion photoproduction}


Inserting the (exact) nucleon current (\ref{eq:Ncurrent3}) into the $s$-channel
term of the photoproduction current (\ref{eq:Mmusplit1}) and using the
splitting (\ref{eq:TXsplit}) of $T$ into its pole and non-pole contributions,
we immediately find
\begin{align}
  M^\mu &= \ket{F} S \tJ_s^\mu  +M^\mu_u +M^\mu_t+M^\mu_c
  \nonumber\\
  &\mbox{}\qquad
  + TG_0\big(M^\mu_u +M^\mu_t +M^\mu_c\big)_\tra~,
  \label{eq:MmuT}
\end{align}
which expresses the final-state interaction in terms of the full $\pi N$ $T$
matrix instead of just its non-polar part $X$. This equation is depicted in
Fig.~\ref{fig:MwithBox}(b).

This reformulation is exact if neither $\tJ^\mu_s$ nor $M^\mu_c$ are
approximated. In practice, however, one employs the approximations
(\ref{eq:JmuPhen}) and (\ref{eq:WTIcontactAmended}) to render the equations
manageable. The corresponding calculations of pion photoproduction are already
underway~\cite{Huang2011}.

The advantage of this reformulation is twofold. First, undesirable numerical
artifacts that may arise from the \textit{non-unique} splitting of $T$ into its
pole part and $X$ can be avoided since $T$ is closer to the actual observables
than $X$. And second, and most importantly, the expression (\ref{eq:MmuT}) only
requires the half on-shell expression of $S\tJ_s^\mu$ for which we can employ
the result (\ref{eq:JsRonshellj}) when we make the approximation $\tJ^\mu_s\to
J^\mu_s$. In actual calculations, one may then use the two coefficient
functions $\kappa_1$ and $\kappa_2$ appearing in the auxiliary currents of
Eqs.~(\ref{eq:auxcurrent}) as fit parameters, which is an excellent
approximation of the dressing effects inherent in the product $SJ_s^\mu$ when
taken half on-shell. This assertion is corroborated by the preliminary
numerical results of Ref.~\cite{Huang2011}.

\section{Discussion and Summary}\label{sec:summary}


Based on the field-theory approach of Haberzettl~\cite{Haberzettl1997}, we have
presented here a formulation of the dressed electromagnetic current of the
nucleon that is microscopically consistent with the reaction mechanisms
inherent in meson photoproduction. The goal was to equivalently rewrite the
original expressions of the full formalism in a manner that retains as much as
possible of its original dynamical structure while at the same time presenting
options for meaningful approximations which in practice are necessary to render
the equations manageable. The consistency requirement, in particular, led to a
novel approximation scheme for pion photoproduction, different from what was
proposed in Ref.~\cite{Haberzettl2006}. The resulting expressions are
summarized diagrammatically in Fig.~\ref{fig:MwithBox} for pion photoproduction
and in Fig.~\ref{fig:NcurrentAll} for the dressed nucleon current.

The full theory is exact. The guiding principle for the construction of the
corresponding equations was the consistent and complete implementation of local
gauge invariance at all levels of the reaction mechanisms in a manner that
lends itself to transparent approximation schemes. In doing so we followed the
basic strategy of Ref.~\cite{Haberzettl2006}, however, with one important and
essential difference. Instead of choosing the optional transverse current
$T^\mu_X$ as zero, as it was done in Ref.~\cite{Haberzettl2006}, we now choose
it so that the resulting expression (\ref{eq:Ncurrent3}) for the dressed
nucleon current exhibits a clean separation of transverse and longitudinal
contributions that makes it straightforward to implement a phenomenological
description of the dressing effects which preserves gauge invariance through
the use of the minimal current $J^\mu_s$ of Eq.~(\ref{eq:JsDefined}).

The phenomenological use of $J^\mu_s$ for the nucleon current makes the
description of pion photoproduction particularly simple when the FSI loop of
the production current $M^\mu$ is written in terms of the full $\pi N$ $T$
matrix (instead of with its non-pole part $X$) because the resulting
$s$-channel term (\ref{eq:JsRonshellj}) then admits a very simple approximation
by utilizing the effective dressing functions $\kappa_1$ and $\kappa_2$ as two
fit parameters.

Another obvious advantage of the present scheme is that for real photons, in
particular, the effective structure of the resulting photoproduction current
remains very close to the full formalism even if the loops over the
five-point-current contributions $U^\mu$ are approximated with the
phenomenological contact current of Eq.~(\ref{eq:WTIcontactAmended}) since for
real photons the longitudinal contributions that make up the structural
difference between the currents $M^\mu$ of Fig.~\ref{fig:MCurrent} and of
Fig.~\ref{fig:MwithBox} are irrelevant.

The approximations discussed here in detail concern replacing the current
$\tilde{J}^\mu_s$ by the minimal current $J^\mu_s$ and the contact current
$M^\mu_c$ of (\ref{eq:McEq2}) by the phenomenological GIP expression
(\ref{eq:WTIcontactAmended}). It should be clear, however, that this still
leaves a formidable self-consistency problem because, as can read off
Fig.~\ref{fig:NcurrentAll}(a), the nucleon current $J^\mu$ also appears in one
of the loops on the right-hand side. In practice, therefore, instead of solving
this self-consistency problem iteratively, one might truncate it at the lowest
level by employing the usual simplified on-shell expression
(\ref{eq:JmuDirPau}) for the current in the loop.

The obvious first application of the present dressing formalism for the nucleon
current is pion photoproduction, of course, since it was the consistency
requirement with this process that inspired the formalism in the first place.
As mentioned, this application is underway already~\cite{Huang2011}, and the
preliminary results obtained so far are very encouraging. In other words, the
present approach is not just formally correct but the approximations suggested
by its formal structure indeed lead to an excellent description of the data.

Other possible applications include any process that may benefit from a
detailed microscopic description of the nucleon current. Obvious candidates are
other meson production processes with both real and virtual photons off the
nucleon, Compton scattering off the nucleon, and $NN$ bremsstrahlung. For
virtual photons, in particular, the present formalism may also be helpful in
extracting the functional behavior of electromagnetic form factors from the
data.

\acknowledgments This work is partly supported by the FFE Grant No.\ 41788390
(COSY-58).


\end{document}